# MatInf – an Extensible Open-Source Solution for Research Digitalisation in Materials Science

Victor Dudarev[0000-0001-7243-9096], Lars Banko[0000-0001-9410-6735], Alfred Ludwig[0000-0003-2802-6774]

Materials Discovery and Interfaces
Ruhr University Bochum
Bochum, Germany
victor.dudarev@rub.de, lars.banko@rub.de, alfred.ludwig@rub.de

## ABSTRACT

Information technology and data science development stimulate transformation in many fields of scientific knowledge. In recent years, a large number of specialized systems for information and knowledge management have been created in materials science. However, the development and deployment of open adaptive systems for research support in materials science based on the acquisition, storage, and processing of different types of information remains unsolved. We propose *MatInf* – an extensible, open-source solution for research digitalisation in materials science based on an adaptive, flexible information management system for heterogeneous data sources. *MatInf* can be easily adapted to any materials science laboratory and is especially useful for collaborative projects between several labs. As an example, we demonstrate its application in high-throughput experimentation.

## KEYWORDS

materials science information infrastructure, materials information management system, inorganic materials information, materials information search

## 1 Introduction

The further development of modern science and technology relies strongly on the collection, curation, processing and use of data and relevant metadata: data-driven science is now more relevant than ever [1]. However, in materials science, despite the growing number of specialized information systems on the properties of materials, there is a lack of flexible open-source independent systems capable of successfully solving the problem of storing and retrieving information about materials and their properties and easily adjusting to support emerging data formats. Successful attempts to develop such fully open-source extendable systems – including both simulated and experimental data - are not known. This might be related to the complexity of the domain and the unavoidable emergence of specific data formats or information structures, which are difficult, if not impossible, to take into account within standalone solutions [2]. Data management is challenging since the system should be flexible to process heterogeneous data formats from proprietary binary data (from measurement devices, e.g., raw files from an X-ray diffraction system) to well-structured and easily parseble XML/JSON/CSV data.

We present MatInf – a highly extensible and customisable platform for materials data management. For the development and implementation, experience from materials science and software engineering were combined. The implementation was supported by extensive usability testing and design reviews of existing data management systems, system analysis and relational data modelling together with UX/UI design and Web development. An analysis of existing Electronic Lab Notebook (ELN) platforms showed that current open-source ELNs: e.g., eLabFTW, openBIS (and other software which are not really ELNs, but are sometimes used in this role: OpenWetWare, JupyterHub) are a good starting point managing scientific data, however, they lack user-defined data types support and customisability and hence are not really able to tackle complex data types properly and perform support and effective search for materials data.

A further motivation for our development was gained through realizing that existing Data Management Systems offered even within NFDI initiatives (e.g., Materials Platform for Data Integration – NFDI-MatWerk; Chemotion – NFDI4Chem; Catalysis-Hub – NFDI4Cat) do not really allow to cover experimental high-throughput materials exploration tasks combined with rising challenges in heterogeneous data and team management together with bridging theoretical and experimental data outcomes within a single information system [3-6].

This article aims at developing a high-level metadata model in materials science, able to be a common ground for materials data, and its implementation in the framework of an open-source system under continuous further development, based on a flexible system of extensible data types (https://gitlab.ruhr-uni-bochum.de/vic/infproject). The current results of the system development and use for representing heterogeneous data on materials properties are demonstrated, and the extension of the system types based on external Web services is briefly discussed. So far, the system is being successfully used in several tenants by geographically distributed collaborating research teams dealing with various tasks and handling different document formats.

## 2 Defining Requirements

Based on our experience in software systems and the materials science domain we formulated requirements for a Research Data Management System (RDMS) capable of materials science data support and specifically to be flexible and powerful enough to tackle changing and emerging data formats used in high-throughput materials exploration tasks. The system could also be used in less demanding research environments and by being able to handle high-throughput workflows it lays also a basis for future autonomous experimentation. The source code is open and the following functional requirements were put forward for the development of the flexible system.

(i) Account support for individual systems (tenants) needs to be implemented. If a new system instance needs to be added, it must be implemented fully independent (separate database, no dependency on proprietary solutions) or be able to share data types and users with other tenants [7].

(ii) A user registration (including e-mail verification) and an authorisation subsystem needs to be developed, including the use of external OpenID Connect-compliant authorisation centres [8]. The system also requires an administrative subsystem for user management, which can be accessed by users with system administrator permissions [9].

For simplicity, the system introduces three predefined user roles.

- *Administrator*: authorised to take any action in the system, including deleting objects and managing users.
- *Power User*: authorised to add and edit objects they have created.
- *User*: has read access rights to protected objects.

If a user does not belong to any of the roles (by default immediately after registration), only access to publicly available information is possible. All information objects in the system have one of three access levels (which is set when the object is created), which determines the visibility of the object for authorised users:

- *Public* – available to everyone, including unauthorised users.
- *Protected* – Available only to authorised users who are assigned to at least the User role.
- *ProtectedNDA* – Available only to authorised users who are assigned to at least the *User* role and have *NDA* (Non-Disclosure Agreement) claim (or in *Administrator* role).
- *Private* – only available to the current user (author).

It should be allowed to load and store different types of objects in a structured way using a user-defined hierarchical classifier. In this way, users can create tree-like rubrics as containers for objects and thus structure the data. It is also possible to establish directional associations between objects, allowing the creation of graphs of interrelated objects.

As the domain of knowledge is defined by materials science, basic type support is required to represent materials objects: material systems, materials and their modifications. The following definitions are introduced:

- *material system* – a set of chemical elements that built up a material, denoted as $S$. Each chemical element is denoted by a unique identifier $e_i$, then the chemical system $S$ is a set $\{e_1, e_2, ..., e_n\}$, in other words, the term "material system" expresses a qualitative composition of a material (e.g., Ni-Ti, Fe-Co-O).
- *material* – a quantified chemical composition, i.e., a set of chemical elements with their general content (if solid solution is considered) or a specific content (in the case of single phase, stoichiometry) that built up a material. Let's denote $n_i$ to represent the quantity or coefficient associated with a particular element $e_i$, then material, designated $C$, is a set of pairs: $C = \{(e_1, n_1), (e_2, n_2), ..., (e_n, n_n)\}$.
- *modification* – e.g., a particular crystal or amorphous structure, form (e.g., thin film, powder, bulk) of a material could be expressed as a string.

The RDMS supports all above-mentioned materials objects. As different types of objects can have an arbitrary set of properties, a mechanism is provided to associate a set of properties of various types with an object, defined at the object type level, with the ability to implement an arbitrary table structure based on the properties that contain the data/metadata according to knowledge/agreements that are defined once one create a type and setup templates. In addition, the import and export of tabular data associated with the object using Excel and the support of Excel templates for tabular data is implemented.

A search interface is developed that supports searches both by built-in object types, including chemical objects (material system, material, modification) and by object properties based on user privileges.

The core of a flexible system is the development of an extensible object type system that not only supports the creation of new object types with their own set of properties based on predefined data types, but also allows the support of new data types by using external, type-supporting services via API. This should perform three main functions: validating data, providing data for import (material systems, materials, properties) into the system to organize search and direct access to data, and data visualisation.

The information system, hereafter denoted as *MatInf*, should provide an API to access data in the system for use in external systems. In addition, visualisation of standard tabular data, and construction of point graphs of dependencies on user-selected axes should be built-in. Other functionality, including the construction of specialized reports, charts and graphs is subject to expansion of the system using external services and APIs.

## 3  Outline of the Architecture of the Extensible Object Structure

In the context of the mentioned requirements, we consider the main implementation aspects of the system. Architecturally it is a modular Web-solution based on ASP.Net Core technology and relational SQL Server database backend. This combination proved its high performance and reliability. Usage of Model-View-Controller design pattern together with high modularity makes the resulting system flexible (see Fig. 1). Abstraction layers, introduced on the top of data and APIs, enable effective data manipulation.

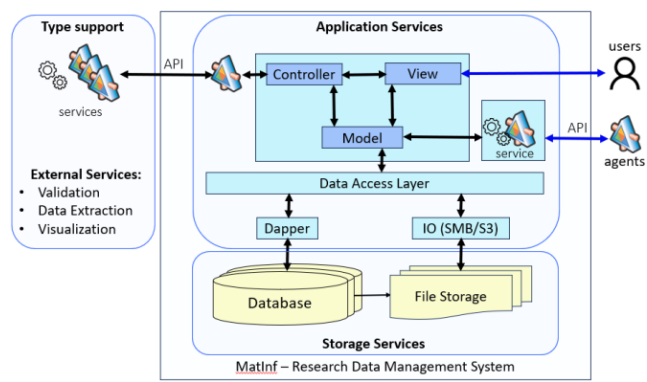

*Figure 1: Schematic of the MatInf architecture.*

The system is based on storage of strongly typed objects in a relational database, where object characteristics (or properties) can be stored in multiple tables. It is important to note that the object definition is introduced by analogy with object-oriented programming, so it is important to define the root object of the hierarchy from which all characteristics are inherited.

### 3.1  Root Class Type

The root class is *ObjectInfo*, which encapsulates properties inherent to all objects, such as a unique object identifier (*ObjectId*); tenant affiliation; date and time of object state creation and modification, and identifier of the user who performed the action; object type identifier (*TypeId*); identifier of object affiliation with the rubric (*RubricId*); sort code (*SortCode* is set if it is necessary to sort objects when outputting in a special way apart from sorting by name, so the *Order By* clause in SQL is "*SortCode, ObjectName*"); object access level (public / protected / protectedNDA / private); optional external object identifier (*ExternalId* – may be necessary for service purposes, e.g., when synchronising the database state with an external resource); object name (mandatory field, must be unique within (*TenantId, TypeId*) pair to prevent wrong naming); object name for building the URL; optional path to the object data file (*ObjectFilePath*); SHA-256 hash of the data file (*ObjectFileHash*, to trace uniqueness of data in the system by building a unique index by the pair (*TenantId, ObjectFileHash*) for filled hashes); optional object description.

All derived classes of objects in the system, such as materials or literary references, are inherited from the *ObjectInfo* type and, in addition to the above-mentioned properties stored in the *ObjectInfo* table, store their state in other system tables (for natively supported types) and extended property tables to store integer, real and string values of object properties (for types supported over native types due to system extensibility).

## 3.2 Materials Classes as Derived Types

Each object belongs to a particular object type, which is described in the relational table *TypeInfo* (referenced by the TypeId attribute). In this context, object types are extensible - a new entry for an object type must be added by referring with the TableName attribute to one of the built-in types (*RubricInfo* - to store hierarchies such as projects/subprojects or organisational structure) for objects: *ObjectInfo*, *Sample* (material system), *Composition* (material), *Reference* (literary reference).

As it is shown in Fig. 2 top-level chemical entities like material system and material are first-class supported, since according to their description in section 2 they are reflected to dedicated tables (*Sample* and *Composition*). This allows the system to interact with corresponding materials objects faster since data structures are optimized at a database level.

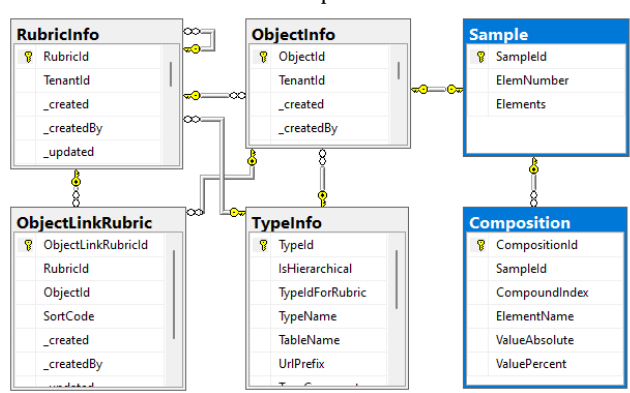

*Figure 2: Sample and Composition tables for materials classes supported by ObjectInfo extension on a database level.*

The same was done with *Reference* table which is connected to *ObjectInfo* by 1-to-1 relation (like *Sample* table, but for the sake of simplicity it is not shown in Fig. 2). Note that each introduced object type may be supported by a typed set of extended properties.

## 3.3 Class Types Extension

Each object can be described by a set of typed properties stored in property value type-dependent tables, as shown in Fig. 3: integer values are stored in *PropertyInt* table, real values in *PropertyFloat* table, string values in *PropertyString* table. Property names are defined by the *PropertyName* attributes, all values are stored in the *Value* attributes of the corresponding *Property\** tables. The build-in separator-property mechanism together with the opportunity to build complex multi-level property names allow to represent hierarchically-arranged properties. This feature is useful when the object has a huge number of parameters that should be hierarchically structured for usability and user convenience. The *Row* attribute allows, if necessary, by specifying the row number, to form a virtual table from the values of the extended properties. This ensures the efficient storage of sparse table data. Obviously, despite of extreme representative power of this mechanism, it should be used in a wise manner: if object number that uses it exceeds 100K and data in this table is not sparse we recommend consider creation of *ObjectInfo*-derived table structures, as it was done with Sample and Composition in Fig. 2.

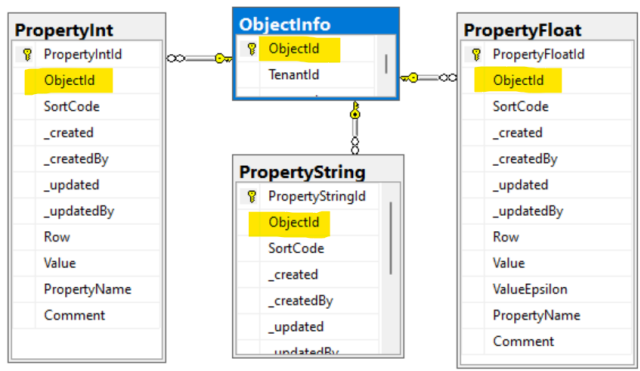

*Figure 3: Extended Object Properties, stored in strongly typed manner.*

## 3.4 Establishing a Dependency Graph

It is often necessary to establish links between objects in the RDMS, e.g., a materials library (consistent collection of materials samples, manufactured under the same conditions [10]) needs to be linked to its measurement results (e.g., measurements of different physico-chemical properties). Thus, importing a compositional analysis document of a materials library not only results in a document of measurement results linked to the library object, but also creates linked objects characterising the quantitative composition of the material in each measurement area (region of a materials library). The directed linking of objects to each other and the assignment of appropriate semantics to links according to the objects' types enables the construction of a graph representation of objects in the materials science domain in question (performed via the *ObjectLinkObject* table).

## 4 API for Flexible Data Type Support

Given that each object can contain a link to a data file (the data obligation for the object type is set in the *TypeInfo.FileRequired* settings), the crucial concerns are data validation, importability into the system and visualisation. Considering the heterogeneity of data in materials science (tabular data in different formats, images, binary files from measurement devices) exchanged by scientists during research, it is impossible to develop a common data schema that would cover *all* possible formats. Therefore, the decision has been made to delegate these tasks to third-party software components, accessible either through late binding via an interface link or using external web services accessible via HTTPS by means of a formalised API. Assignment of the task of working with data types unknown at the moment of system design to third-party services with their response schema formalisation allows to work with files of different formats with equal success, since external software components are responsible for their analysis and return documents of standardised format.

## 4.1 Data Validation

In addition to mandatory validation for uniqueness of stored files for each object type (at *TypeInfo* table level), one can specify an additional validation method by modifying *TypeInfo.ValidationSchema* value. Options are either a built-in validator supporting *IFileValidator* interface, which instance is created dynamically using Reflection (using "type:" prefix), or external validator available via REST API (using "https:" prefix). In the second case, the document for validation is sent to the address specified in the settings in the POST request body, and the service returns serialized in a JSON object of *TypeValidatorResult* type with validation results described by three properties:

- *Code – int* type, error code: 0 – validation succeeds; != 0 – validation fails.
- *Message – string* type, error message if validation fails (null if validation succeeds).
- *Warning – string* type, warning on successful validation (otherwise null). Message that may cause an error in the future (in case of rules strictification).

So, if the document validation is successful the response returned is {"Code": 0, "Message": null, "Warning": null}. If the validation failed, {"Code": 500, "Message": "Error description", "Warning": null}. The service specification is available in OpenAPI format, which simplifies the development of similar services to support new object types.

## 4.2 Data Import

For data search in the system and indexing of uploaded documents, special support for data import schema is introduced at type level, supporting the addition of chemical objects as well as extended properties when a document is uploaded. This support, architecturally similar to validation, can be implemented either at the embedded object level with the help of predefined programming interfaces or at the level of external web services accessible via REST APIs. A complete specification for the import data formats is not possible here, however, the formats from the DTO object specifications are available in the source code of the system (see *TypeValidationLibrary* project).

## 4.3 Data Visualisation

The RDMS has built-in support for visualising graphics in standard formats and tabular data, with the functionality to sort them and build point plots on selected axes. Although support for visualisation of materials libraries and their properties is additionally implemented (Fig. 4), it is evident that this is not sufficient for flexible handling of arbitrary data formats, especially binary formats, the support of which may introduce a separate challenge. Therefore, visualisation of type data can be configured using an external web application accessible via the HTTP(S) protocol.

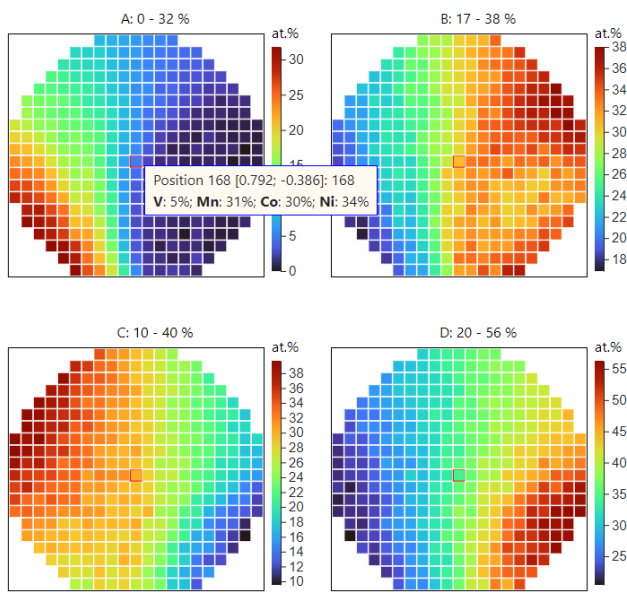

*Figure 4: Example of a materials library visualisation showing color-coded compositional information of a quaternary system.*

An external rendering address can be specified for the object type, capable of accepting a POST request from a system with data in the body of the request. Having received the data, the external system independently renders the document and is responsible for interaction with the user.

## 5   Data Search

The data search is materials-oriented and offers flexible searching options to find the required objects. The described flexible type system plays a major role in this, which makes a decisive contribution to the information content of the database when importing heterogeneous documents. The user interface of the search subsystem is presented in Fig.5. The search form is dynamically adjustable allowing the user to add properties to search in an underlying type-aware way: numbers could be searched within a specified range and strings using substring search. Moreover, search is facilitated by automatic suggestions for parameter names of user selected types by looking up database content. Ability to search for materials entities using object properties together with implemented persistent search URLs that can be shared to reference some results make search one of the crucial advantages in the current software system. It should be noted, that the search is carried out taking into account the current security context (depends on the authorised user and his membership to roles) and the level of availability of objects in the system.

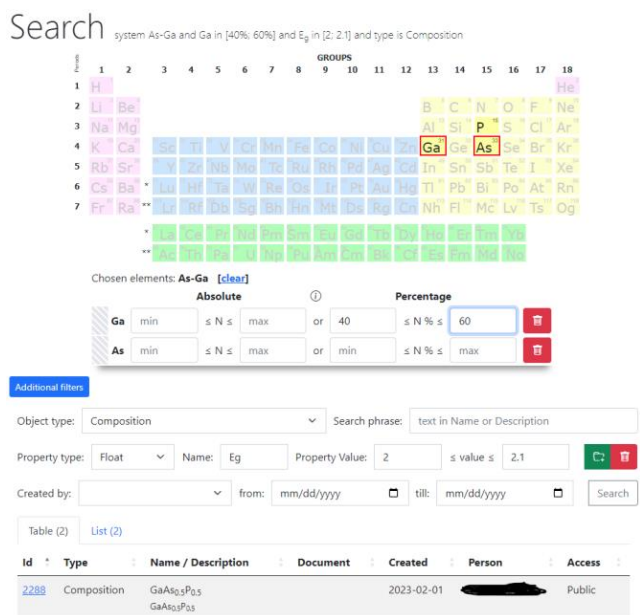

*Figure 5: Exemplary Search Results for Composition containing Ga (in range 40-60%) and As (any content) with Bandgap in range 2-2.1 eV.*

## 6 Conclusion

The open-source RDMS is under continuous development and belongs to the class of Research Data Management Systems for materials science, since it provides support for materials science object types. The extensibility of the system by adding and supporting validation, importing and mapping data from added types through the use of APIs gives the system considerable versatility, allowing flexible work with heterogeneous data types. Support of multiple tenants allows not only isolation of system instances, but also fast addition of new instances, which gives the possibility to provide the developed system as SaaS (Software as a Service) solution for different workgroups. External interface and object type system options allow for flexible customisation. Further development steps are aimed at enhancing the customisation of individual tenants and implementing an API for direct input of data into the system e.g., from measurement devices or custom data sources, which will enable automated research data collection and further analysis, including the use of third-party software systems communicating via API.

### ACKNOWLEDGMENTS

This research was financially supported by the Deutsche Forschungsgemeinschaft (DFG, German Research Foundation) Project-ID 388390466-TRR 247 (subproject INF).